\documentclass[twocolumn,showpacs,preprintnumbers,amsmath,amssymb]{revtex4}

\usepackage[dvips]{graphicx}
\usepackage{dcolumn}
\usepackage{bm}

\begin{document}

\preprint{HMI/Notbohm I}

\title{One- and Two-Triplon Spectra of a Cuprate Ladder}

\author{S. Notbohm$^{1\!,2}$, P. Ribeiro$^3$, B. Lake$^{1\!,4}$, D.A. Tennant$^{1\!,4}$,
 K.P. Schmidt$^5$, G.S. Uhrig$^6$, C. Hess$^3$, R. Klingeler$^3$, G.
Behr$^3$, B. B\"{u}chner$^3$, M. Reehuis$^1$, R.I. Bewley$^7$, C.D.
Frost$^7$, P. Manuel$^7$, and R.S. Eccleston$^8$}
\address{
 $^1$Hahn-Meitner-Institut Berlin, Glienicker Stra\ss e 100, 14109 Berlin, Germany\\
 $^2$School of Physics and Astronomy,\ University of St.\,Andrews, St.\,Andrews,\ Fife, KY16 9SS, U.K.\\
 $^3$Leibniz-Institut for Solid State and Material Research,\ IFW-Dresden,\ 01171,\ Germany \\
 $^4$Institut f\"{u}r Festk\"{o}rperphysik, Technische Universit\"{a}t Berlin,\ Hardenbergstr. 36,\
10623 Berlin,\ Germany \\
 $^5$Institute of Theorectical Physics,\ \'{E}cole Polytechnique F\'{e}d\'{e}rale de Lausanne,\ 1015 Lausanne,\ Switzerland\\
 $^6$Lehrstuhl f\"{u}r Theoretische Physik I,\ Universit\"{a}t Dortmund,\ 44221 Dortmund,\ Germany\\
 $^7$ISIS Facility,\ Rutherford Appleton Laboratory,\ Chilton,\ Didcot OX11 OQX,\ U.K. \\
 $^8$Materials and Engineering Research Institute,\ Sheffield Hallam University,\ Howard Street,\ Sheffield,\ S1 1WB,\ U.K.
 }%

\date{\today}

\begin{abstract}
We have performed inelastic neutron scattering on the near ideal
spin-ladder compound La$_4$Sr$_{10}$Cu$_{24}$O$_{41}$, which will
serve as a model system for the investigation of doped ladders. A
key feature in the analysis was the separation of one-triplon and
two-triplon scattering contributions due to an antiphase
rung modulation. This enabled extraction of the one-triplon
dispersion which has a spin gap of $26.4\!\pm\!0.3$\,meV, and for
the first time two-triplon scattering is clearly observed. The
exchange constants are determined by fitting the data to be
\textit{J$_{leg}$}=186\,meV and \textit{J$_{rung}$}=124\,meV
along the leg and the rung respectively. In addition to the
surprisingly large difference between \textit{J$_{leg}$} and
\textit{J$_{rung}$} a substantial ring exchange of
\textit{J$_{cyc}$}=31\,meV was confirmed.
\end{abstract}

\maketitle

The two-leg ladder, consisting of two spin-1/2 chains coupled by a
Heisenberg interaction, can be thought of as a strip, two sites wide,
cut out of a 2D square lattice antiferromagnet. The immediate effect
of coupling the chains is to confine pairs of spin-1/2 spinons, and
to generate a gapped spin singlet ground state. Short ranged singlet
spin pairings on the rungs ${1/\sqrt{2} \{|\uparrow\downarrow\rangle
- |\downarrow\uparrow\rangle\}}$ dominate the ground state, and in
an inelastic neutron scattering measurement a triplet of well-defined spin-1 magnons, known as
triplon \cite{schmidt03} excitations, rather than a spinon pair continuum will appear.
Further, strong higher-order ground state fluctuations should be
manifest as multi-triplon continua. The spin pairings are particularly
important as they provide a mechanism for the formation of
charged hole pairs when doped, giving rise to condensed charge
density wave or superconducting states
\cite{dagotto92,rice93,gopalan94,tsunetsugu95,troyer96}. Cuprate
spin ladders share many features of the 2D high-$T_c$ materials
(pseudo-gap, non-Fermi liquid behavior)
\cite{takahashi97,dagotto99,tranquada04}, and have similar coupling
parameters. Their excitation spectra extend to several hundred meV
and the development of time-of-flight spectroscopy at spallation
neutron sources allows access to the full spectra resolved in
wave-vector and energy for the first time. Motivated by this, we
have measured the one- and two-triplon spectra of a cuprate
ladder in a Mott-Hubbard insulating (MHI) phase -
La$_4$Sr$_{10}$Cu$_{24}$O$_{41}$- using inelastic neutron scattering
(INS). A full description of the dynamics is achieved using the
continuous unitary transformation (CUT) method. Electrostatic
environments of the oxygen tunneling result in strongly dissimilar
rung and leg exchanges, and this combined with a substantial cyclic
charge fluctuations render the magnetic state of the system near
quantum critical (gapless).

The geometry of cuprate ladders in La$_4$Sr$_{10}$Cu$_{24}$O$_{41}$
is depicted in the upper part of
Fig.\,1. Strong anti-ferromagnetic super-exchange
occurs through the 180$^\circ$ Cu-O-Cu bonds with very weak
exchange between ladders via the 90$^\circ$ bonds. An approximate
electronic Hamiltonian for the cuprates is the one-band Hubbard
model
\begin{equation}
H=-\sum_{\langle i,j\rangle \sigma=\uparrow,\downarrow} t_{i,j}\,
(c^{+}_{i,\sigma} c_{j,\sigma} + H.c.) + U \sum_{i}
n_{i,\uparrow}n_{i,\downarrow}
\end{equation}
with on-site Coulomb energy $U\!\sim\!3.5$ eV, inter-site hopping
between nearest neighbor pairs of $t\!\sim\!0.3$ eV, and
$\kappa\!\equiv\!t/U\!\sim\!\frac{1}{12}$ \cite{yang06}. Charge fluctuations such as cyclic
hopping processes, (suppressed in the $\kappa\rightarrow \!0$ limit),
are reflected in modifications to the Heisenberg spin Hamiltonian
with the cyclic hopping contributing a four spin interaction term.
In fact computations in the more realistic 3-band Hubbard model
\cite{mueller-hartmann02} (which include O2p and Cu3d orbitals)
indicate only short range Heisenberg and four spin interactions are
important with diagonal coupling across the CuO square plaquette and
further neighbor interactions negligible. The resulting spin
Hamiltonian is
\begin{equation}
H=J_{rung}\sum_i {\bf S}_{i,1} \cdot {\bf S}_{i,2}+
J_{leg}\sum_{i,\tau} {\bf S}_{i,\tau} \cdot {\bf S}_{i+1,\tau}+
H_{cyc} 
\end{equation}
\begin{eqnarray*}
H_{cyc}=J_{cyc}\sum_{plaquettes}[\,({\bf S}_{i,1} \cdot {\bf
S}_{i+1,1})({\bf S}_{i,2} \cdot {\bf S}_{i+1,2}) +\\
({\bf S}_{i,1}\cdot {\bf S}_{i,2})({\bf S}_{i+1,1} \cdot {\bf
S}_{i+1,2})- ({\bf S}_{i,1} \cdot {\bf S}_{i+1,2})({\bf S}_{i+1,1}
\cdot {\bf S}_{i,2}) ],
\end{eqnarray*}
The anticipated dispersion relation for the spin-1 triplon
states is plotted in the lower part of Fig.\,1. For the ladder
the relative ratios of the gaps at wave-vectors along the leg
direction $Q_c\!=0,\,0.25,\,0.5$ (in units of $\pi/2c)$ are
modified substantially by the strength of $J_{cyc}$.

\begin{figure}
\includegraphics[width=0.8\linewidth]{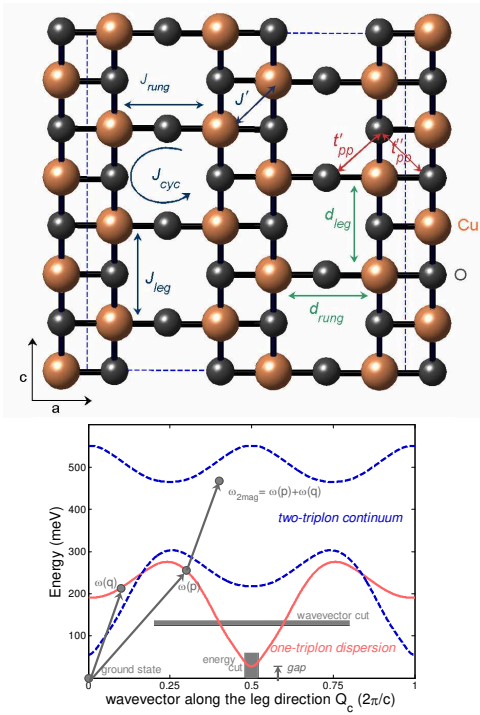}
\caption{\label{fig:fig_1} (Color online) Upper part: Arrangement of the ladder
 planes in La$_{4}$Sr$_{10}$Cu$_{24}$O$_{41}$.
 Big spheres represent the copper ions, small spheres oxygen
 atoms.
 The dashed rectangle displays the unit cell in the (010) crystallographic
 plane for the \textit{a} axis and half of the unit cell in the
 \textit{c} axis. Lower part: $S$=1 excitation spectrum.
 There is a gap from the $S$=0 ground state to the $S$=1 triplon band (red solid curve).
 The two-triplon continuum states bounded by the two blue dashed curves can be
 constructed via the scattering of the two triplons illustrated by the arrows.
 A typical wave-vector cut (horizontal)
 and energy cut (vertical) are indicated.}
\end{figure}

Also shown are the boundaries of the two-triplon
continuum. The intensity of the two-triplon spectrum, wave-vector
and energy distribution is strongly dependent on the triplon
interactions and exact composition of the ground state. Importantly,
a bound spin-1 state is predicted below the two-triplon continuum.
The four spin exchange term frustrates the formation of this bound
mode lowering the binding energy and for coupling strengths of order
$x_{cyc}=J_{cyc}/J_{rung}\sim0.25$ will disrupt binding completely.

A remarkable property of the ladder geometry is that the sectors with even and
odd triplon number do not mix because of different parity w.r.t. 
reflection about the centerline of the ladder. Hence even and odd 
contributions can be Fourier decomposed by the rung coupling, 
and the neutron scattering matrix elements can be
expressed as a product of a neutron structure factor (NSF) dependent
only on the wave-vector along the leg direction $q_c$, multiplied by
a rung wave-vector $q_a$ modulation.  The one- and two-triplon
spectra are in anti-phase with rung modulations $(1-\cos (q.a))$ and
$(1+\cos (q.a))$ respectively  \cite{schmidt05}, and this can be used to separate them
in a neutron scattering experiment.

Here we investigate La$_{4}$Sr$_{10}$Cu$_{24}$O$_{41}$: Although 
intrinsically hole-doped, X-ray absorption spectroscopy
reveals that in $\rm La_{x}(Sr,Ca)_{14-x}Cu_{24}O_{41}$ with , 
residual holes are located in the more electronegative
chains \cite{nuecker00}, leaving ladders undoped
\cite{carter96,tennant04}. No triplon-hole scattering is observed
in the heat conductivity \cite{hess01,hess04}.

\begin{figure}[t]
\includegraphics[width=0.75\linewidth]{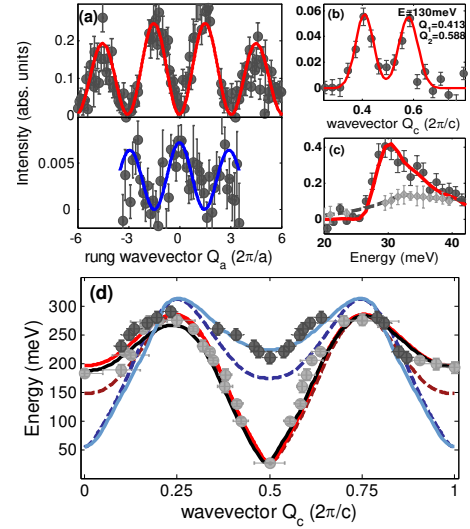}
\caption{\label{fig:fig_2} (Color) (a) Cut along the rung wave-vector shows the intensity modulation
 Upper part: for energy of $30\!\pm\!5$\,meV and leg wave-vector of $0.45\!\pm\!0.05$ in the
rung direction for the one-triplon scattering. Lower part: for energy of $300\!\pm\!10$\,meV and
leg wave-vector of $0.25\!\pm\!0.05$  for two-triplon.
Solid line described in text.
 (b) Typical wave-vector cut through one-triplon along \textit{Q$_{c}$} for energy of $130\!\pm\!5$\,meV.
 The solid line is a Gaussian fit to extract peak position.
 (c) Energy cut for $0.48\!<\!\textit{Q$_{c}$}\!<\!0.52$  to
determine the energy gap. Light grey points are the corresponding cut for a
hole-doped ladder.
 (d) Experimental points from neutron data. Theoretical one-triplon dispersion 
 curves shown in solid red and black. Black curve
 includes interladder coupling. Theoretical lower boundary of one-triplon scattering
  shown in solid blue. Dashed lines show respective scattering without cyclic exchange.}
\end{figure}

Large $\rm La_{4}Sr_{10}Cu_{24}O_{41}$ single crystals were prepared
using the `\,Travelling Solvent Floating Zone'-method at 9 bar
oxygen pressure. Inelastic neutron scattering measurements were
performed using the MAPS spectrometer at ISIS, Rutherford Appleton
Laboratory, U.K. Three crystals of total mass 23\,g were co-aligned
with resulting mosaic spread of 0.4$^\circ$ in the
\textit{b}-\textit{c} plane and 5.5$^\circ$ on the
\textit{a}-\textit{c} plane. The samples were mounted with the (0kl)
reciprocal lattice plane horizontal, and the \textit{c} axis (leg
direction) perpendicular to the incident neutron wave-vector
\textit{k$_{i}$} in a cryostat giving a measurement temperature of
10\,K. The wave-vectors are labeled \textit{Q$_{c}$} along
the leg and \textit{Q$_{a}$} along the rung. The magnetic
cross-section was normalized using incoherent nuclear scattering
from a vanadium standard. Data were collected for incident energies
\textit{E$_{i}$}=\,75.55,\,363.7 and 606.2\,meV. The energy runs
were monochromated using a Fermi chopper at 300\,Hz for the low
energy run and at 600\,Hz for the other runs.

\begin{figure}[t]

\includegraphics[width=\linewidth]{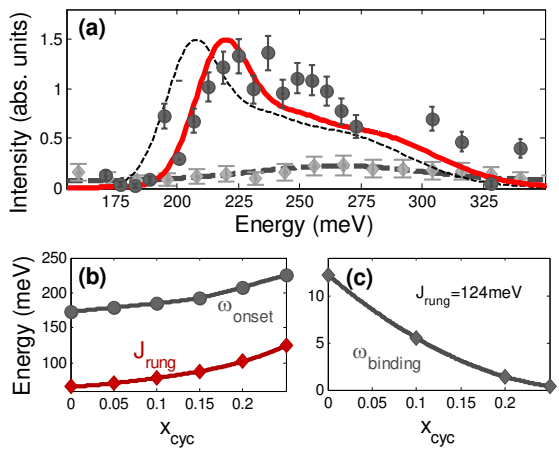}
 \caption{\label{fig:fig_3} (Color online) (a) Energy cut for two-triplon at $Q_c\!=\!0.5\!\pm\!0.1$.
 Solid line is two-triplon model. Intensities had to be increased by 15$\,\%$ to match data.
 Light grey points are the corresponding cut for a hole-doped ladder. 
 Thin dashed line is the effect of absent cyclic exchange for fixed $J_{rung}$.
 (b) Relationship of $\omega_{onset}$ and $J_{rung}$ w.r.t. $x_{cyc}$. For an onset energy of 225\,meV
 $J_{rung}=124$\,meV and $x_{cyc}=0.25$ are needed (see text). $J_{cyc}=x_{cyc}\,J_{rung}$. (c) Frustration
 of bound mode with increasing cyclic exchange. $\omega_{binding}$ difference between
 $\omega_{onset}$ and energy of bound mode.}
\end{figure}

Contributions mainly from one-triplon or two-triplon scattering were
separated using the antiphase rung modulation. The measured
modulation is shown in Fig.\,2\,(a) with the one-triplon modulation
fitted to $(1-\cos(q \cdot a))$ and the two-triplon modulation to
$(1+\cos(q \cdot a))$, with corrections for the anisotropic magnetic
form factor of copper \cite{shamoto93} applied to the fitted
functions. One-triplon data used in subsequent analysis was taken at
the maxima between $1\!<\!|\textit{Q$_a$}|\!<\!2$, see in
Fig.\,2\,(a). For the two-triplon data, two regions of maximum
two-triplon intensity $0\!<\!|\textit{Q$_a$}|\!<\!0.5$ and
$2.65\!<\!|\textit{Q$_a$}|\!<\!3.35$ were added taking magnetic form
factor corrections fully into account.

The non-magnetic background correction for the one-triplon signal was
deduced via interpolation of the intensity found at energies above
and below the one-triplon dispersion. The background subtracted data
is displayed in Fig.\,4\,(a). For the two-triplon scattering, the
background was deduced from the scattering observed at
$\textit{Q$_c$}=0,\,1,\,2$ where the magnetic signal is
expected to be weak. The result is shown in Fig.\,4\,(c).

The one-triplon dispersion was determined by extracting positions and
intensities of measured peaks by least-squares fitting of Gaussians.
A typical cut is shown in Fig.\,2\,(b), this gives \textit{Q$_c$} at
energy of $130\!\pm\!5$\,meV indicated in Fig.\,1 by the horizontal
slice. These along with data points from different cuts are
displayed in Fig.\,2\,(d). Heisenberg couplings alone are unable to
account for the energy gaps of $197,286,26.4$ meV, at wavevectors
$0,\,0.25,\,0.5$ respectively. Our computations using the CUT method
\cite{schmidt05} reveal that a substantial four-spin component
\textit{J$_{cyc}$}=\,31\,meV and
Heisenberg exchange strengths along the legs
\textit{J$_{leg}$}=\,186\,meV, and rungs
\textit{J$_{rung}$}=\,124\,meV that are very different \cite{schmidt2005} are required to explain this. Computing the entire
dispersion using these values (solid red line) results in excellent
agreement with the data {\cite{RPA}}. The lower boundary of the
two-triplon scattering was extracted in the same manner and the data
points along with the computed lower boundary of the two-triplon
continuum (blue curve) are plotted in Fig.\,2\,(d). Also shown
is the computed one-triplon (and two-triplon) dispersion in the
absence of cyclic exchange (dashed curves). One can see the
necessity of introduction of the cyclic exchange to describe the
data within its error bars.

The CUT method is also used to compute the NSF which agrees very
well with the measured scattering, see Fig.\,4(b) and (d). A
complete simulation of detector coverage and multicrystal mount is
included with the detailed angular mosaics of each crystal, measured
via neutron Laue, accounted for using a Monte Carlo integration. The
uneven intensity profile of the data with wavevector is due to a
combination of detector coverage and crystalline mosaic. Fig.\,2(c)
shows an energy cut at $Q_c=\!0.5$ (vertical slice in Fig.\,1,
wavevector range $0.48\!<\!|\textit{Q$_{c}$}|\!<\!0.52$). The
computed NSF including instrumental corrections is shown as the red
line and matches the data. Note the fitted gap is
$26.4\!\pm\!0.3$\,meV which is lower than found in previous
measurements \cite{eccleston96,azuma94}.

The effect of cyclic exchange on the two triplon binding is illustrated
in Fig.\,3. The data points in Fig.\,3\,(a) show a two-triplon
wave-vector cut of $Q_c=\,0.5\!\pm\!0.1$; the red solid line is the
same cut performed through the model with the determined exchange
parameters. Fig.\,3\,(b) illustrates the dependence of the continuum
onset energy $\omega_{onset}$ (measured as 224\,meV), with $J_{rung}$ determined to
match the one triplon gaps at $Q_c=0.25,\,0.5$, on cyclic
exchange ratio $x_{cyc} \equiv J_{cyc}/J_{rung}$. Keeping $J_{rung}$
fixed and assuming no cyclic exchange a bound mode 12\,meV below $\omega_{onset}$ would
appear, shown in Fig.\,3\,(c) $\omega_{binding}=12$\,meV. Such a
shift is indicated by the black dashed line in Fig.\,3\,(a) and
would be detectable within our resolution limits confirming we are
on the binding threshold.

\begin{figure}[t]
\includegraphics[width=\linewidth]{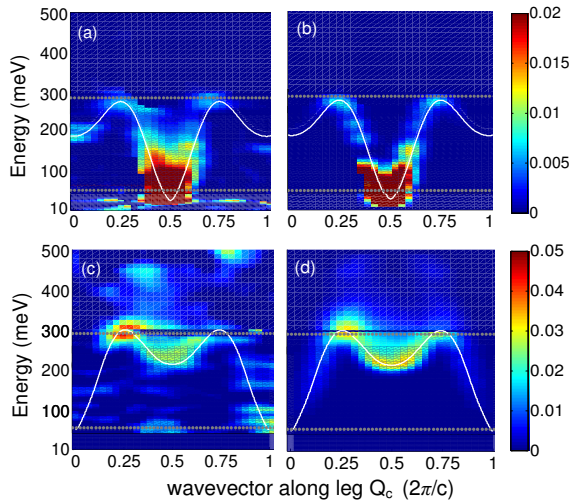}
\caption{\label{fig:fig_4}(Color) (a) Measured one triplon data with
non-magnetic background subtracted. The white curve gives the
theoretically calculated one triplon dispersion curve. (b) Calculated
one-triplon scattering with corrections for instrumental resolution
and magnetic form factor. (c) Measured two triplon data with
nonmagnetic background subtracted. The white curve gives the lower
boundary of the two-triplon dispersion curve. (d) Calculated
two-triplon scattering.}
\end{figure}

Figs.\,4\,(c)\,and\,(d) show that nearly all the $S=1$ two-triplon
scattering weight is concentrated near the lower continuum boundary,
and this is reproduced very well by calculation. Van Hove
singularities in the density of states of phonon assisted IR
absorption measurements indicate that $S=0$ bound states survive as
they are more stable in quantum antiferromagnets \cite{schmidt05, windt01,nunner02}. The CUT method also predicts the relative one and two triplon neutron
scattering intensity scales. The scale factors applied to the model
above are within $15\%$ of the computations, i.e. within experimental
accuracy.


The large difference between rung and leg exchange implies
stronger hopping along the legs $t_{leg}>t_{rung}$ despite a
longer pathway (\textit{d$_{leg}$}=\,3.96\AA,
\textit{d$_{rung}$}=\,3.84\AA). Contributions from hopping between
O2p orbitals \cite{eskes93} depend acutely on electrostatic
environment \cite{mizuno98}. For ladders, Fig.\,1, O2p hopping along 
the leg,
\textit{t$^{\prime\prime}_{pp}$} is enhanced by two neighboring
Cu$^{2+}$ ions compared to the rung,
\textit{t$^{\prime}_{pp}$} which has one Cu$^{2+}$ neighbor
\cite{mizuno98} accounting for the stronger $t_{leg}$.
Further, admixture of Cu4s states has been proposed
to enhance $t_{leg}$ hopping (by 35\,\%)
\cite{arai97,mueller98} compared to $t_{rung}$. Comparison of our
values of \textit{J$_{leg}$}, \textit{J$_{rung}$}, and
\textit{J$_{cyc}$} with theoretical calculation should lead to a
more refined model for cuprate electronic structure.

Approximate parameters for (1), $t_{leg}\!=\!0.42$, $t_{rung}\!=\!0.34$, $U\!=\!3.72$\:eV,
follow from low order perturbation theory \cite{Hubbard}, but recourse to 3-band
computations is necessary for high accuracy. To account for the CDW states shown at low
temperatures in hole-doped ladders
Sr$_{14-x}$Ca$_{x}$Cu$_{24}$O$_{41}$ \cite{nagata98} electrostatic
interactions $\sum_{i,a=L,R} V_{leg} n_{i,a} n_{i+1,a} + V_{rung}
n_{i,L} n_{i,R}$ have been introduced in an extended Hubbard model \cite{wu03}.
Our MAPS measurements on the $x\!=\!11.5$ doped ladder ($\rho
\propto T$ above $T_{CDW} \sim 80K$) which has a modest doping of
$5\%$ holes on the ladder, show the one-triplon spectra to
split into multiple branches and for the gap to develop into
a pseudogap (see 2(c)). Further the boundaries of the two-triplon
spectra undergo dramatic changes and the lower boundary at
$Q_c\!=\!0.5$ collapses (see Fig.\,3(a)). Understanding the dramatic
evolution of the spin correlations from the MHI to doped
metallic ladder is a challenging problem, but is necessary to
bridge the gap between understanding INS
from insulating quantum magnets versus strongly correlated metals
such as high-$T_c$.

In conclusion, our high resolution INS measurements establish the
full one- and two-triplon spectra in the MHI
cuprate ladder, and also the accuracy of the continuous unitary
transformation method. Computation of the magnetic spectra in doped
ladders within the framework of an extended Hubbard model is urgently needed.

We thank T.G. Perring, H.-J. Mikeska, A.L. L\"{a}uchli, P.
Schubert-Bischoff and S.-L. Drechsler for support.


\begin{references}

\bibitem{schmidt03}  K.P. Schmidt {\it et al.}, Phys. Rev. Lett. {\bf 90}, 227204 (2003).

\bibitem{dagotto92}  E. Dagotto {\it et al.}, Phys. Rev. B {\bf 45},
5744 (1992).

\bibitem{rice93}  T.M. Rice {\it et al.}, Europhys. Lett. {\bf 23},
445 (1993).

\bibitem{gopalan94} S. Gopalan, T.M. Rice, and M. Sigrist, Phys.
Rev. B {\bf 49}, 8901 (1994).

\bibitem{tsunetsugu95} H. Tsunetsugu, M. Troyer, and T.M. Rice,
Phys. Rev. B {\bf 51}, 16456 (1995).

\bibitem{troyer96} M. Troyer, H. Tsunetsugu, and T.M. Rice, Phys.
Rev. B {\bf 53}, 251 (1996).

\bibitem{takahashi97}  M. Takahashi {\it et al.}, Physica B {\bf 237-238},
112 (1997).

\bibitem{dagotto99} E. Dagotto, Rep. Prog. Phys {\bf 62}, 1525
(1999).

\bibitem{tranquada04} J.M. Tranquada {\it et al.}, Nature {\bf 429},
534 (2004).

\bibitem{yang06} K.-Y.Yang {\it et al.}, cond-mat$\backslash$0603423.

\bibitem{mueller-hartmann02} E. M\"{u}ller-Hartmann and A. Reischl, Eur. Phys J. B {\bf 28}, 173 (2001)


\bibitem{schmidt05}  K.P. Schmidt, and G.S. Uhrig, Mod. Phys. Lett. B {\bf 19}, 1179 (2005).

\bibitem{nuecker00} N. N\"{u}cker {\it et al.}, Phys. Rev. B {\bf 62},
14384 (2000).

\bibitem{carter96}  S.A. Carter {\it et al.}, Phys. Rev. Lett. {\bf 77},
1378 (1996).

\bibitem{tennant04}  D.A. Tennant {\it et al.},  {\it unpublished}.

\bibitem{hess01} C. Hess {\it et al.}, Phys. Rev. B {\bf 64},
184305 (2001).

\bibitem{hess04} C. Hess {\it et al.}, Phys. Rev. Lett. {\bf 93},
027005 (2004).

\bibitem{shamoto93}  S. Shamoto {\it et al.}, Phys. Rev. B {\bf 48}
13817 (1993).


\bibitem{schmidt2005} K.P. Schmidt {\it et al.}, Phys. Rev. B {\bf 72}, 094419 (2005).

\bibitem{RPA} We also explored interladder effects within the RPA. The black line in Fig.\,2(b) is
the dispersion computed including an interchain exchange
\textit{J'}=\,36\,meV. The interladder coupling does not have a
large effect on the disperion and is undetermined with experimental
accuracy.

\bibitem{eccleston96}  R.S. Eccleston {\it et al.}, Phys. Rev. B {\bf 53},
R14721 (1996).

\bibitem{azuma94}  M. Azuma {\it et al.}, Phys. Rev. Lett. {\bf 73},
3463 (1994).

\bibitem{windt01}  M. Windt {\it et al.}, Phys. Rev. Lett. {\bf 87}, 127002 (2001).

\bibitem{nunner02} T. Nunner {\it et al.}, Phys. Rev. B {\bf 66}, 180404(R) (2002).

\bibitem{eskes93}  H. Eskes, and J.H. Jefferson, Phys. Rev. B {\bf 48},
9788 (1993).

\bibitem{mizuno98}  Y. Mizuno {\it et al.}, Phys. Rev. B {\bf 58},
R14713 (1998).

\bibitem{arai97}  M. Arai, and H. Tsunetsugu, Phys. Rev. B {\bf 56},
R4305 (1997).

\bibitem{mueller98}  T.F.A M\"{u}ller {\it et al.}, Phys. Rev. B {\bf 57}
R12655 (1998).

\bibitem{Hubbard} Parameters to $0(\kappa^4): $ $J_{leg}\!=\!4t_{leg}^2/U+O(\kappa^4)$,\\
$J_{rung}\!=\!4t_{rung}^2/U+O(\kappa^4)$,
$J_{cyc}\!=\!80t_{leg}^2t_{rung}^2/U^3$; A. Rusydi {\it et al.}, Phys. Rev. Lett {\bf 97}, 
016403 (2006).

\bibitem{nagata98} T. Nagata {\it et al.}, Phys. Rev. Lett. {\bf 81},
1090 (1998).

\bibitem{wu03} C.J. Wu, L.W. Vincent, and E. Fradkin , Phys. Rev. B {\bf 68}
115104 (2003).













\end{references}
\end{document}